\documentclass[twoside,10pt,a4paper]{newFNLstyle}
\usepackage{amssymb}

\usepackage{graphics}
\usepackage{cite}

\begin{document}

\volnumpagesyear{2}{4}{R189--R203}{2002}
\dates{21 May 2002}{21 November 2002}{29 December 2002}

\title{PLAYING PRISONER'S DILEMMA WITH QUANTUM RULES
\thanks{Fluctuation and Noise Letters, {\it Special Issue on Game Theory and Evolutionary Processes:
Order from Disorder --- The Role of Noise in Creative Processes},
edited by D. Abbott, P.C.W. Davies and C.R. Shalizi.}}

\authorsone{JIANGFENG DU}
\affiliationone{Department of Modern Physics, \\ University of
Science and Technology of China, \\ Hefei, 230027, People's
Republic of China\\ and \\
Department of Physics, \\ National University of Singapore, Lower Fent Ridge, \\ Singapore 119260} %

\authorstwo{XIAODONG XU, HUI LI, XIANYI ZHOU, and RONGDIAN HAN}
\affiliationtwo{Department of Modern Physics, \\ University of
Science and Technology of China, \\ Hefei, 230027, People's
Republic of China}

\mailingone{ djf@ustc.edu.cn }

\maketitle

\markboth{J. Du et al.}{Playing Prisoner's Dilemma with Quantum
Rules}

\pagestyle{myheadings} 


\begin{abstract}
Quantum game theory is a recently developing field of physical research. In
this paper, we investigate quantum games in a systematic way. With the
famous instance of the Prisoner's Dilemma, we present the fascinating
properties of quantum games in different conditions, \textit{i.e.} different
number of the players, different strategic space of the players and
different amount of the entanglement involved.
\end{abstract}

\section{Introduction}

Game theory is a distinct and interdisciplinary approach to the
study of human behavior. The foundation of modern game theory
can be traced back to the mathematician John Von Neumann who, in
collaboration with the
mathematical economist Oskar Morgenstern, wrote the mile stone book \textit{%
Theory of Games and Economic Behavior}\cite{rf1}. Game
theory has since become one of the most important and useful tools for
a wide range of research from the economics, social science to
biological evolution\cite{rf2}.

However, game theory is now being developed in a completely new way. It is extended into the quantum realm by physicists interested in quantum
information theory\cite{rf3}. In quantum game theory, the players play the game
abiding by quantum rules. Therefore, the primary results of quantum games
are very different from those of their classical counterparts\cite%
{rf4,rf4-1,rf5,rf6,rf7,rf8,rf9,rf10,rf11, rf12,rf13}. For example,
in an otherwise fair zero-sum coin toss game, a quantum player can
always win against his classical opponent if he adopts quantum
strategies\cite{rf4,rf4-1}. In some other original dilemma games,
problems can be resolved by playing with quantum
rules\cite{rf5,rf6,rf7,rf11,rf13}. Besides the theoretical
research on quantum games, a quantum game has been experimentally
realized on a NMR quantum computer\cite{rf9,rf14}.

In this paper, we investigate quantum games in a systematic way. For the
particular case of the quantum Prisoner's Dilemma, we investigate this
quantum game in different conditions. These conditions differ in the number
of players, the strategic space and the role of entanglement in the quantum
game. It is shown that the properties of the quantum game can change
fascinatingly with the variations of the game's conditions.

\section{Quantum Games}

In the remaining part of this paper, we take a detailed
investigation of quantum games with the particular case of the
Prisoner's Dilemma. Firstly, we present the two-player Prisoner's
Dilemma which has been presented in \cite{rf5}. Secondly, we
investigate the three-player Prisoner's Dilemma. In the following
discussions the organization is as follows: (i) the classical
version of the game, (ii) the quantization scheme, and (iii) the
investigation of the game with different strategic spaces.

\subsection{Two-player prisoner's dilemma}

\begin{table}[t]
\caption{The payoff table for the Prisoners' Dilemma.}
\label{table}
\begin{center}
\begin{tabular}{ccc}
\hline & Bob: $C$ & Bob: $D$ \\ \hline
Alice: $C$ & $\left( 3,3\right) $ & $\left( 0,5\right) $ \\
Alice: $D$ & $\left( 5,0\right) $ & $\left( 1,1\right) $ \\ \hline
\end{tabular}%
\end{center}
\end{table}

The publication of \textit{Theory of Games and Economic Behavior} was a
particularly important step in the development of game theory. But in some
ways, Tucker's proposal of the problem of the Prisoners' Dilemma was even more
important. This problem, which can be stated in one page, could be the
most influential one in the social sciences in the later half of the
twentieth century. The name of the Prisoner's Dilemma arises from
the following scenario: two burglars, Alice and Bob are caught by the
police and are interrogated in separate cells, without no communication
between them. Unfortunately, the police lacks enough admissible evidence to
get a jury to convict. The chief inspector now makes the following offer to
each prisoner: If one of them confess to the robbery, but the other does
not, then the former will get unit reward of 5 units and the latter will get nothing. If
both of them confess, then each get 1 unit as a reward. If neither of them
confess, then each will get payoff 3. Since confession means a
\textquotedblleft defect\textquotedblright\ strategy and no confession
means \textquotedblleft cooperate\textquotedblright\ with the other player,
the classical strategies of the players are thus denoted by
\textquotedblleft $D$\textquotedblright\ and \textquotedblleft $C$%
\textquotedblright , respectively. Table \ref{table} indicates the
payoffs of Alice and Bob according to their strategies.

From Table \ref{table}, we see that $D$ is the dominant strategy of the game.%
\textit{\ }Since the players are rational and care only about their
individual payoffs, both of them will resort to the dominant strategy $D$
and get payoff $1$. In terms of game theory, $(D,D)$ is a dominant strategy
equilibrium. However, this dominant strategy equilibrium is inferior to the
Pareto Optimal $(C,C)$, which yields payoff 3 to each players. This is the
catch of the Prisoner's Dilemma.

\subsubsection{Quantization scheme}

\begin{figure}[t]
\centering{\includegraphics{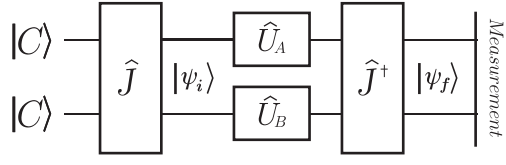}} \caption{The setup of
two-player quantum games.} \label{Fig1}
\end{figure}

Recently, this famous game got a new twist: It is studied in the
quantum world by physicists\cite{rf5,rf6}. By allowing the players
to adopt quantum strategies, it is interesting to find that the
original dilemma in the classical version of this game could be
removed. The physical model of this quantum game is illustrated in
Fig \ref{Fig1}. Different from the classical game, each player has
a qubit and can manipulate it independently (locally) in the
quantum version of this game. The quantum formulation proceeds by
assigning the possible outcomes of the classical strategies $C$
and $D$ the two
basis vectors of a qubit, denoted by%
\begin{equation}
\left\vert C\right\rangle =\left(
\begin{array}{c}
1 \\
0%
\end{array}%
\right) ,\left\vert D\right\rangle =\left(
\begin{array}{c}
0 \\
1%
\end{array}%
\right) .
\end{equation}%
The gate%
\begin{equation}
\hat{J}=\exp \left\{ i\gamma \hat{D}\otimes \hat{D}/2\right\} ,
\end{equation}%
with $0\leqslant \gamma \leqslant \pi /2$, can be considered as a gate which
produce entanglement between the two qubits. The game started from the pure
state $\left\vert CC\right\rangle $. After passing through the gate $\hat{J}$%
, the game's initial state is%
\begin{equation}
\left\vert \psi _{i}\right\rangle =\widehat{J}\left\vert CC\right\rangle
=\cos \frac{\gamma }{2}\left\vert CC\right\rangle +i\sin \frac{\gamma }{2}%
\left\vert DD\right\rangle .
\end{equation}%
Since the entropy (entanglement) of $\left\vert \psi _{i}\right\rangle $ is%
\begin{equation}
S=-\sin ^{2}\frac{\gamma }{2}\ln \sin ^{2}\frac{\gamma }{2}-\cos ^{2}\frac{%
\gamma }{2}\ln \cos ^{2}\frac{\gamma }{2},
\end{equation}%
the parameter $\gamma $ can be reasonably considered as a measure of the
game's entanglement.

After the initial state was produced, each player apply a unitary operation
on his/her individual qubit. Later on, the game's state goes through $\hat{J}%
^{\dag }$ and the final state is $\left\vert \psi _{f}\right\rangle $.
According to the corresponding entry of the payoff table (Table \ref{table}%
), the explicit expressions of both player's payoff functions can be written
as follows:

\begin{eqnarray}
\$_{A} &=&3P_{CC}+1P_{DD}+5P_{DC}+0P_{CD},  \nonumber \\
\$_{B} &=&3P_{CC}+1P_{DD}+0P_{DC}+5P_{CD},
\label{payoff-function}
\end{eqnarray}%
where $\$_{A}$ ($\$_{B}$) represents Alice's (Bob's) payoff and
$P_{\sigma \sigma ^{\prime }}=\left\vert \left\langle \sigma
\sigma ^{\prime }\right\vert \left. \psi _{f}\right\rangle
\right\vert ^{2} $ is the probability that the final state will
collapse into $\left\vert \sigma \sigma ^{\prime }\right\rangle $.
At the end of the game, each player will get a reward according
the payoff function.

In the following subsections, we investigate this quantum game with
different strategic spaces. It is interesting to find that the game's
property does not necessarily become better with the extension of the
strategic space.

\subsubsection{Restricted strategic space}

In this section, we will focus on the restricted strategic space situation,
\textit{i.e.} a two parameter strategy set which is a subset of the whole
unitary space\cite{rf5}. The explicit expression of the operator is given by%
\begin{equation}
U(\theta ,\phi )=\left(
\begin{array}{cc}
e^{i\phi }\cos \frac{\theta }{2} & \sin \frac{\theta }{2} \\
-\sin \frac{\theta }{2} & e^{-i\phi }\cos \frac{\theta }{2}%
\end{array}%
\right),
\end{equation}%
where $0\leqslant \theta \leqslant \pi $ and $0\leqslant \phi \leqslant \pi
/2$. We can see that%
\begin{equation}
U(0,0)=\left(
\begin{array}{cc}
1 & 0 \\
0 & 1%
\end{array}%
\right)
\end{equation}%
is the identity operator and%
\begin{equation}
U(\pi ,0)=\left(
\begin{array}{cc}
0 & 1 \\
-1 & 0%
\end{array}%
\right)
\end{equation}%
is somehow equivalent to a bit-flip operator. The former corresponds to the
classical \textquotedblleft \textit{cooperation}\textquotedblright\ strategy
and the latter to \textquotedblleft \textit{defect}\textquotedblright .

This situation has been investigated in details by Eisert \textit{et al}.%
\cite{rf5}. Here, we present the main results of their work: (i) For a
separable game with $\gamma =0$, there exists a pair of quantum strategies $%
\hat{D}\otimes \hat{D}$, which is the Nash equilibrium and yields payoff $%
\left( 1,1\right) $. Indeed, this quantum game behaves \textquotedblleft
classically\textquotedblright , \textit{i.e.} the Nash equilibrium for the
game and the final payoffs for the players are the same as in the classical
game. So the separable game does not display any features which go beyond
the classical game. (ii) For a maximally entangled quantum game with $\gamma
=\pi /2$, there exists a pair of strategies $\hat{Q}\otimes \hat{Q}$, which
is a Nash equilibrium and yields payoff $\left( 3,3\right) $, having the
property to be the \textit{Pareto optimal}. Therefore the dilemma that
exists in the classical game is removed.

In the quantum game, we can see that in the decision-making step the player
has means of communication with each other, \textit{i.e.} no one has any information
about which strategy the other player will adopt. This is the same as in
classical game. Hence, it is natural to ask why the dilemma game shows such
a fascinating property in quantum game? The answer is \textit{entanglement},
the key to the quantum information and quantum computation\cite{rf15,rf16}.
Although there is no communication between the two players, the two qubits
are entangled, and therefore one player's local action on his qubit will affect
the state of the other. Entanglement plays as a contract of the game.

In the Eisert \textit{et al}.'s scheme, the dilemma was removed when the
game's state is maximal entangled. It is also interesting to investigate the
game's behavior when the amount of entanglement varies. In one of our
previous works\cite{rf6}, we find that there exist two thresholds of the
game's entanglement, $\gamma _{th1}=\arcsin \sqrt{1/5}$ and $\gamma
_{th2}=\arcsin \sqrt{2/5}$. In different domains of entanglement, the
quantum game shows different properties. For $0\leqslant \gamma \leqslant
\gamma _{th1}$ , the quantum game behaves classically, \textit{i.e.} the
Nash equilibrium of the game is $\hat{D}\otimes \hat{D}$ and the final
payoff for the players are both $1$, which are the same as the classical
version of this game. Hence,the quantum game does not display any features
which go beyond the classical game for the small amount of entanglement of
the game's state. For $\gamma _{th1}\leqslant 0\leqslant \gamma _{th2}$, the
games shows some novel features which has no classical analog. In this
domain, $\hat{D}\otimes \hat{D}$ is no longer Nash equilibrium of the game.
However, there are two new Nash Equilibria, $\hat{D}\otimes \hat{Q}$ and $%
\hat{Q}\otimes \hat{D}$. The payoff to the one who resorts to strategy $\hat{%
D}$ is $5\cos ^{2}\gamma $ and to the other adopting $\hat{Q}$ is $5\sin
^{2}\gamma $. Since $5\cos ^{2}\gamma \geqslant 5\sin ^{2}\gamma $ for $%
\gamma _{th1}\leqslant \gamma \leqslant \gamma _{th2}$, the one
choosing strategy $\hat{D}$ is better rewarded. Note that the
physical structure of the game is symmetric with respect to the
interchange of the two players. However, both Nash equilibrium
$\hat{D}\otimes \hat{Q}$ and $\hat{Q}\otimes \hat{D}$ cause the
the unfairness of the game. We think that there are two reasons
for the asymmetry situation: (i) Since the definition of Nash
equilibrium allows multiple Nash Equilibria to coexist, the
solutions may be degenerate. Therefore the definition itself
allows the possibility of such an asymmetry. This situation is
similar to the spontaneous symmetry breaking; (ii) If we consider
the two Nash Equilibria as a whole, they are fully equivalent and
the game remains symmetric. But finally, the two players have to
choose one from the two equilibria. This also causes the asymmetry
of the game. For $\gamma _{th2}\leqslant \gamma \leqslant \pi /2$,
the game shows exciting features. A novel Nash equilibrium
$\hat{Q}\otimes \hat{Q}$ arises with
$\$_{A}(\hat{Q},\hat{Q})=\$_{B}(\hat{Q},\hat{Q})=3$, which
satisfies the property of Pareto Optimal. Therefore, as long as
the the amount of entanglement exceeds a certain threshold, the
dilemma can be removed.

\begin{figure}[t]
\centering{\includegraphics{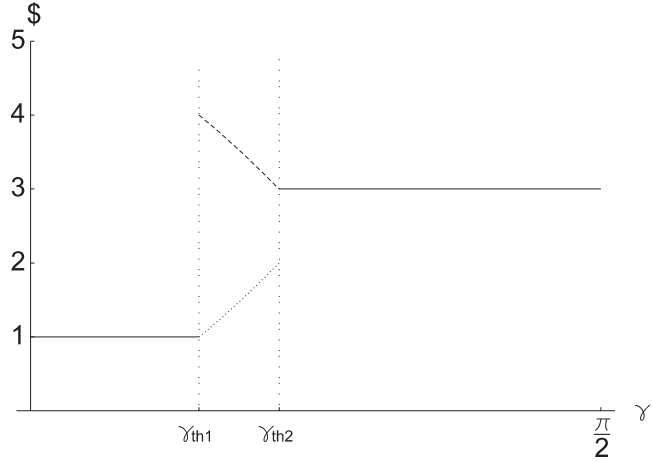}} \caption{The expected
payoff for Alice as a function of the measure of the parameter
$\protect\gamma $\ when both players resort to Nash Equilibrium.
For $\protect\gamma _{th1}<\protect\gamma <\protect\gamma _{th2}$,
the dash
line and the dot line represent Alice's payoff when the Nash Equilibrium is $%
\hat{D}\otimes \hat{Q}$\ and $\hat{Q}\otimes \hat{D}$\
respectively.} \label{Fig2}
\end{figure}

Fig \ref{Fig2} illustrates Alice's payoff as a function of the parameter $%
\gamma $ when both players resort to Nash equilibrium. From this figure, we
observe that the game's property depends discontinuously on the amount of
the entanglement. This discontinuity can be considered as entanglement
correlated phase transition, \textit{i.e.} the game can be considered to lie
in three different phases. For $0\leqslant \gamma \leqslant \gamma _{th1}$,
the game displays no advantage over classical game. So this domain can be
considered as the classical phase. For $\gamma _{th1}\leqslant \gamma
\leqslant \gamma _{th2}$, there are two Nash Equilibria for the game, both
of which yield asymmetric payoffs to the players. This domain can be
considered as the transitional phase from classical to quantum. For $\gamma
_{th2}\leqslant \gamma \leqslant \pi /2$, a novel Nash equilibrium $\hat{Q}%
\otimes \hat{Q}$ appears with payoff $\left( 3,3\right) $. This
strategic profile has the property to be Pareto Optimal and hence
the dilemma disappears, and this domain can be considered as the
quantum phase.

One interesting thing should be pointed out is that if we change
the numerical values in the payoff table (Table \ref{table}), Fig
\ref{Fig2} will varies interestingly. If these numerical values
satisfy some particular condition, the transition phase in which
the game has two asymmetric Nash equilibria will disappear.
Furthermore, the phase transition exhibit interesting variation
with respect to the change of the numerical values in the payoff
matrix, so does the property of the game. For different numerical
values, the game may or may not have a transition phase, or even
the classical and quantum phases can overlap and form a new phase,
the
coexistence phase. The detailed presentation can be found in \cite{rf18}%
.

\subsubsection{General quantum operations}

In a recent Letter, it was pointed out that restricting the strategic space
of the players can not reflect any reasonable physical constraint because
the set is not closed under composition\cite{rf17}. The observation is that
any operation of the restricted strategic space can consist of $\left\{ \hat{%
I},\hat{\sigma}_{y},\hat{\sigma}_{z}\right\} $ with certain coefficients. $%
\hat{I}$ equivalent to $\hat{C}$ operation, $\hat{\sigma}_{y}$ to
$\hat{D}$ and $\hat{\sigma}_{z}$ to $\hat{Q}$. Note that
$\hat{\sigma}_{y}$ is an optimal strategy counter to $\hat{I}$.
And similarly $\hat{\sigma}_{z}$ is the ideal counter-strategy to
$\hat{\sigma}_{y}$. However, the best reply to strategy
$\hat{\sigma}_{z}$, which is $\hat{\sigma}_{x}$, is not included
in this restricted strategic space. Thus, the dilemma situation
could be solved by applying $\hat{Q}\otimes \hat{Q}$. The general
case is that the player should be permitted free choice of any
unitary operations. If so, the operation $\hat{\sigma}_{x}$
counter to $\hat{\sigma}_{z}$ is permitted. The interesting
observation is that $\hat{I}$ is the ideal counter-strategy if
one's opponent plays $\hat{\sigma}_{x}$. Then $\left\{ \hat{I},\hat{\sigma}%
_{x},\hat{\sigma}_{y},\hat{\sigma}_{z}\right\} $ forms a
inter-restricted cycle, and therefore if the amount of
entanglement of the game's state is maximal, there is no pure-strategy Nash equilibrium in this quantum game\cite{rf17,rf17-1}.
However, the game remains to have mixed Nash
equilibria\cite{rf17-1}.

As we have seen in the preceding section, properties of quantum games change
fascinatingly when the amount of the game's entanglement varies. So,
assuming that the player could choose any strategy from the complete set of all
local unitary operations, it is then natural to investigate whether there
exists pure strategy Nash equilibrium for this game when the game is not
maximally entangled. Here we show that as long as the game's
entanglement is below a certain boundary\cite{rf6}, the game has infinite number of
pure Nash equilibria. While for entanglement beyond that boundary, the game
behaves the same as in the maximally entangled case.

\begin{figure}[t]
\centering{\includegraphics{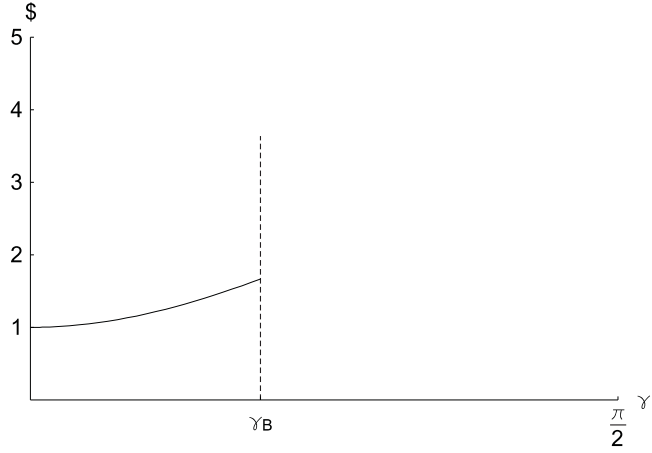}} \caption{The expected
payoff for Alice as a function of the measure of the parameter
$\protect\gamma $\ when both players resort to Nash Equilibrium.}
\label{Fig3}
\end{figure}

\bigskip The general form of $2\times 2$ unitary matrix can be represented
by Pauli matrices as following:%
\begin{equation}
\hat{U}(w,x,y,z)=w\hat{I}+ix\cdot \hat{\sigma}_{x}+iy\cdot \hat{\sigma}%
_{y}+iz\cdot \hat{\sigma}_{z},  \label{unitary operation}
\end{equation}%
where all the coefficients $w$, $x$, $y$ and $z$ are real and satisfy the
normalization condition%
\begin{equation}
w^{2}+x^{2}+y^{2}+z^{2}=1.
\end{equation}%
We plot Alice's payoff as a function of the parameter $\gamma $
when both players resort to Nash equilibrium in Fig \ref{Fig3}.
$\gamma _{B}=\arcsin \sqrt{1/3}$ is the boundary of the game's
entanglement. For $0\leqslant \gamma \leqslant \gamma _{B}$ , we
find that there are infinite number of
Nash equilibrium for any determined value of $\gamma $. But as long as the $%
\gamma $ is given, no matter which profile Nash equilibrium the players
choose, the payoffs for both players are determined and are the same, which
are $\$_{A}=$ $\$_{B}=1+2\sin ^{2}\gamma $. It shows that the payoff of each
player is a monotonous increasing function of $\gamma $. But if the
entanglement of the game's state exceeds the boundary $\gamma _{B}$, the
game will have no pure strategy Nash equilibrium. So we have shown that if
the strategic space of the players is all of $SU\left( 2\right) $, the
entanglement could still enhance the property of this quantum game.

\subsection{Multiplayer quantum games}

The effect of \textquotedblleft two's company, three's a
crowd\textquotedblright\ is quite familiar in physical world\cite{rf10}.
Complex phenomenon tends to emerge in multipartite systems. Hence, to
investigate multiplayer quantum games in multi-qubit system will be more
interesting and significant. Recently, quantum games with more than two
players was firstly investigated and such games can exhibit certain forms of
pure quantum equilibrium that have no analog in classical games, or even in
two player quantum games\cite{rf11,rf19}. In the following discussion, we
investigate multiplayer quantum games with the particular case of the
three-player Prisoner's Dilemma. Since the structure of the game is
symmetric, and for more explicit expression, our investigation focus on the
symmetric solution of the game. In this case, we will show that the quantum
game can display miscellaneous qualities under different conditions. At
first, let's extend the two-player Prisoner's Dilemma to the three-player
case.

\subsubsection{Three-player prisoner's dilemma}

The scenario of the three player Prisoners' Dilemma is similar to
the two-player situation\cite{rf1}. Besides Alice and Bob, a third
player, Colin joins this game. They are picked
up by the police and interrogated in separate cells without a
chance to communicate with each other. For the purpose of this
game, it makes no difference whether or not Alice, Bob or Colin
actually committed the crime. The players are told the same thing:
If they all choose strategy $D$ (defect), each of them will get
payoff $1$; if the players all resort to strategy $C$ (cooperate),
each of them will get payoff $3$; if one of the players choose $D$
but the other two do not, $5$ is payoff for the former and $2$ for
the latter two; if one of the players choose $C$ while the other
two adopt $D$, $0$ is payoff for the former and $4$ for the latter
two.

\begin{figure}[t]
\centering{\includegraphics{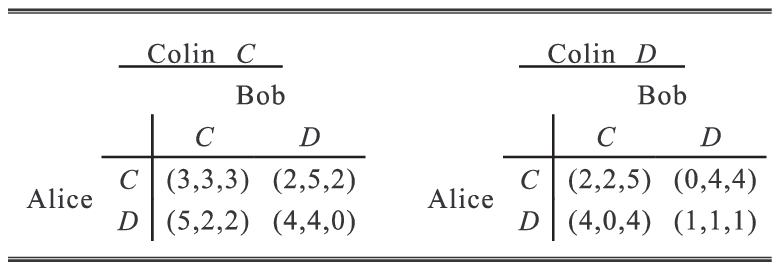}} \caption{The payoff Table
of the 3-player Prisoners' Dilemma. The first entry in the
parenthesis denotes the payoff of Alice, the second number denotes
the payoff of Bob, and the third number denotes the payoff of
Colin.} \label{Fig4}
\end{figure}

Fig \ref{Fig4} indicates the payoffs of the three players
depending on their decisions. The game is symmetric for the three
players, and the strategy $D$ dominates strategy $C$ for all of
them. Since the selfish
players all choose $D$ as optimal strategy, the unique Nash equilibrium is $%
(D,D,D)$ with payoff $(1,1,1)$. This is a Pareto inferior outcome, since $%
(C,C,C)$ with payoffs $(3,3,3)$ would be better for all three players.
Optimizing the outcome for a subsystem will in general not optimize the
outcome for the system as a whole. This situation is the very catch of the
dilemma and the same as the two-player version of this game.

\subsubsection{Quantization scheme}

\begin{figure}[t]
\centering{\includegraphics{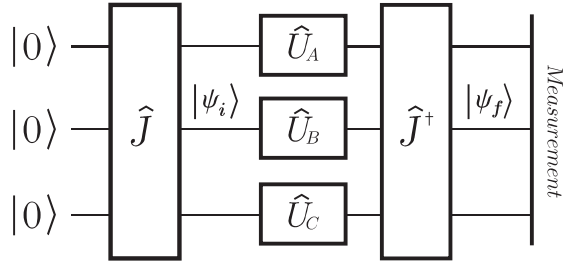}} \caption{The setup of
three-player quantum games.} \label{Fig5}
\end{figure}

Our physical model for quantizing this game is similar as in
\cite{rf11}
--- see Fig \ref{Fig5}. Just like the quantization scheme of the two-player
Prisoner's Dilemma, here we send each player a two state system or a qubit
and they can locally manipulate their individual qubit. The possible
outcomes of the classical strategies \textquotedblleft \textit{Cooperate}%
\textquotedblright\ and \textit{\textquotedblleft Defect}\textquotedblright\
are assigned to two basis vector%
\begin{equation}
\left\vert 0\right\rangle =\left(
\begin{array}{c}
1 \\
0%
\end{array}%
\right) ,\left\vert 1\right\rangle =\left(
\begin{array}{c}
0 \\
1%
\end{array}%
\right)
\end{equation}%
in the Hilbert space. In the procedure of the game, its state is described
by a vector in the tensor product space which is spanned by the eight
classical basis $\left\vert \sigma \sigma ^{^{\prime }}\sigma ^{^{\prime
\prime }}\right\rangle $ ( $\sigma ,\sigma ^{^{\prime }},\sigma ^{^{\prime
\prime }}\in \left\{ 0,1\right\} $), where the first, second and third
entries belonging to Alice, Bob and Colin, respectively. At the beginning of
the game the game's state is $\left\vert 000\right\rangle $. After the
unitary transformation $J$, the initial state of the game is%
\begin{equation}
\left\vert \psi _{i}\right\rangle =\hat{J}\left\vert 000\right\rangle ,
\end{equation}%
where%
\begin{equation}
\hat{J}=\exp \left\{ i\frac{\gamma }{2}\sigma _{x}\otimes \sigma _{x}\otimes
\sigma _{x}\right\} ,
\end{equation}%
with $0\leqslant \gamma \leqslant \pi /2$, is the entangling gate of the
game and is known to all of the players. Strategic move of Alice (Bob or
Colin) is denoted by unitary operator $\hat{U}_{A}$ ($\hat{U}_{B}$ or $\hat{U%
}_{C}$), which are chosen from a certain strategic space $S$. Since the
strategic moves of different players are independent, one player's operator
just operates on his individual qubit. After the operations of the players,
the final state of the game prior to the measurement is give by%
\begin{eqnarray}
\left\vert \psi _{f}\right\rangle &=&\left\vert \psi _{f}\left( \hat{U}_{A},%
\hat{U}_{B},\hat{U}_{C}\right) \right\rangle  \nonumber \\
&=&\hat{J}^{\dag }\left( \hat{U}_{A}\otimes \hat{U}_{B}\otimes \hat{U}%
_{C}\right) \hat{J}\left\vert 000\right\rangle .  \label{eq 2}
\end{eqnarray}%
Hence, the final state of the game can be represented by density matrix
\[
\rho _{f}=\left\vert \psi _{f}\right\rangle \left\langle \psi
_{f}\right\vert .
\]%
Here, we will use the density matrix to get the final payoff for the
players. This final state go forward to the subsequent measurement
instrument and the players can get a reward according to their individual
payoff operators. The payoff operators can be directly given from the
corresponding entries of payoff matrix. For example, the payoff operator of
Alice can be written as
\begin{eqnarray}
\hat{\$}_{A} &=&5\left\vert 100\right\rangle \left\langle 100\right\vert
+4\left( \left\vert 110\right\rangle \left\langle 110\right\vert +\left\vert
101\right\rangle \left\langle 101\right\vert \right)  \nonumber \\
&&+3\left\vert 000\right\rangle \left\langle 000\right\vert +2\left(
\left\vert 001\right\rangle \left\langle 001\right\vert +\left\vert
010\right\rangle \left\langle 010\right\vert \right)  \nonumber \\
&&+1\left\vert 111\right\rangle \left\langle 111\right\vert +0\left\vert
011\right\rangle \left\langle 011\right\vert .  \label{threePDpayofffunction}
\end{eqnarray}%
Hence the expectation value of Alice's payoff is given by%
\begin{equation}
\$_{A}\left( U_{A},U_{B},U_{C}\right) =tr\left[ \hat{\$}_{A}\rho _{f}\right]
.  \label{eq 4}
\end{equation}%
Since the game is symmetric for three players, the payoff
functions of Bob and Colin can also be obtained directly from the
same analyzing together with the payoff matrix (see Fig
\ref{Fig4}).

\subsubsection{Two-parameter strategic set}

We start investigation of three-player Prisoner's Dilemma with restricted
strategic space, \textit{i.e.} the strategic space is a two-parameter set%
\cite{rf19}. The matrix representation of the corresponding operators is
taken to be
\begin{equation}
\hat{U}\left( \theta ,\varphi \right) =\left(
\begin{array}{cc}
\cos \theta /2 & e^{i\varphi }\sin \theta /2 \\
-e^{-i\varphi }\sin \theta /2 & \cos \theta /2%
\end{array}%
\right)  \label{eq 1}
\end{equation}%
with $0\leqslant \theta \leqslant \pi $ and $0\leqslant \varphi \leqslant
\pi /2$. To be specific, $\hat{U}\left( 0,0\right) $ is the identity
operator $\hat{I}$ which corresponds to \textquotedblleft \textit{Cooperate}%
\textquotedblright , and $\hat{U}\left( \pi ,\pi /2\right) =i\sigma _{x}$,
which is equivalent to the bit-flipping operator, corresponds to
\textquotedblleft \textit{Defect}\textquotedblright . Therefore $\hat{J}$
commutes with any operator formed from $i\sigma _{x}$ and $\hat{I}$ acting
on different qubits, and this guarantees that the classical Prisoners'
Dilemma is faithfully entailed in the quantum game.

If there is no entanglement (for $\gamma =0$), the game is separable,
\textit{i.e.} at each instance the state of the game is separable. We find
that any strategy profile formed from $\hat{U}\left( \pi ,\pi /2\right)
=i\sigma _{x}$ and $\hat{U}\left( \pi ,0\right) =i\sigma _{y}$ is Nash
equilibrium. However this property of multiple equilibria is a trivial one.
For any profile of Nash equilibrium of the separable game, because $i\sigma
_{x}\left\vert 0\right\rangle =i\left\vert 1\right\rangle $ and $i\sigma
_{y}\left\vert 0\right\rangle =-\left\vert 1\right\rangle $, the final state
$\left\vert \psi _{f}\right\rangle =-\left( -i\right) ^{n}\left\vert
1\right\rangle \left\vert 1\right\rangle \left\vert 1\right\rangle $, where $%
n$ denotes the number of players who adopts $\hat{U}\left( \pi /2,\pi
\right) =i\sigma _{x}$. According to the payoff functions in Eq. (\ref%
{threePDpayofffunction}), each player receives payoff $1$. Hence in this
case $i\sigma _{x}$ and $i\sigma _{y}$ have the same effect to the payoffs.
And in this sense, all the Nash Equilibria are equivalent to the classical
strategy profile $(D,D,D)$. Indeed, the separable game does not exceed the
classical game.

Although unentangled games is trivial, the behavior of maximally entangled
game is fascinating and surprising. The profile $i\sigma _{x}\otimes i\sigma
_{x}\otimes i\sigma _{x}$ is no longer the Nash equilibrium. However, a new
Nash equilibrium, $i\sigma _{y}\otimes i\sigma _{y}\otimes i\sigma _{y}$ $%
\left( i\sigma _{y}=U(\pi ,0)\right) $, emerges with payoffs%
\begin{equation}
\$_{A}\left( i\sigma _{y},i\sigma _{y},i\sigma _{y}\right) =\$_{B}\left(
i\sigma _{y},i\sigma _{y},i\sigma _{y}\right) =\$_{C}\left( i\sigma
_{y},i\sigma _{y},i\sigma _{y}\right) =3.
\end{equation}%
Indeed, for $\gamma =\pi /2$,%
\begin{equation}
\$_{A}\left( \hat{U}\left( \theta ,\varphi \right) ,i\sigma _{y},i\sigma
_{y}\right) =\left( 2+2\cos 2\varphi \right) \sin ^{2}\frac{\theta }{2}%
\leqslant 3=\$_{A}\left( i\sigma _{y},i\sigma _{y},i\sigma _{y}\right)
\end{equation}%
for all $\theta \in \left[ 0,\pi \right] $ and $\varphi \in \left[ 0,\pi /2%
\right] $. Analogously%
\begin{eqnarray}
\$_{B}\left( i\sigma _{y},\hat{U}\left( \theta ,\varphi \right) ,i\sigma
_{y}\right)  &\leqslant &\$_{B}\left( i\sigma _{y},i\sigma _{y},i\sigma
_{y}\right) ,  \nonumber \\
\$_{C}\left( i\sigma _{y},i\sigma _{y},\hat{U}\left( \theta ,\varphi \right)
\right)  &\leqslant &\$_{C}\left( i\sigma _{y},i\sigma _{y},i\sigma
_{y}\right) .
\end{eqnarray}%
Hence, no player can improve his individual payoff by unilaterally
deviating from the strategy $i\sigma _{y}$, \textit{i.e.} $\left(
i\sigma _{y},i\sigma _{y},i\sigma _{y}\right) $ is a Nash
equilibrium. It is interesting to see that the payoffs for the
players are $\$_{A}=\$_{B}=\$_{C}=3$, which are the best payoffs
that retain the symmetry of the game. Thus the strategy profile
$(i\sigma _{y},i\sigma _{y},i\sigma _{y})$ has the property of
Pareto Optimal, \textit{i.e.}\ no player can increase his payoff
without lessening the payoff of the other players by deviating
from this pair of strategies. Therefore by allowing the players to
adopt quantum strategies, the dilemma that exists in the classical
game is completely removed when the game is maximally entangled.

In the above paragraph, we have considered the maximally entangled game. In
this case, an novel Nash equilibrium $\left( i\sigma _{y},i\sigma
_{y},i\sigma _{y}\right) $ emerges, which has the property of Pareto
optimal. Since the key role of entanglement in quantum information, it will
be interesting to investigate whether this strategy profile is still Nash
equilibrium when the game is not maximal entangled. And if it is, how the
property of the game changes with the variations of the entanglement when
the players each resort to $i\sigma _{y}=U(\pi ,0)$. The surprising thing is
that $i\sigma _{y}\otimes i\sigma _{y}\otimes i\sigma _{y}$ is always a Nash
equilibrium for any $\gamma \in \left( 0,\pi /2\right) $. The proof that
this pair of strategy is still Nash equilibrium runs as follows. Assume Bob
and Colin adopt $i\sigma _{y}$ as their strategies, the payoff function of
Alice respect to her strategy $\hat{U}\left( \theta ,\varphi \right) $ is%
\begin{eqnarray}
\$_{A}\left( \hat{U}\left( \theta ,\varphi \right) ,i\sigma _{y},i\sigma
_{y}\right) &=&\left( 1+2\cos ^{2}\varphi \sin ^{2}\gamma \right) \sin ^{2}%
\frac{\theta }{2}  \nonumber \\
&\leqslant &1+2\sin ^{2}\gamma =\$_{A}\left( i\sigma _{y},i\sigma
_{y},i\sigma _{y}\right) .
\end{eqnarray}%
So, $i\sigma _{y}$ is her best reply to the other players. Since
the game is symmetric, the same holds for Bob and Colin.
Therefore, no matter what the amount of the game's entanglement
is, $i\sigma _{y}\otimes i\sigma _{y}\otimes i\sigma _{y}$ is
always a Nash equilibrium for the game. It is fascinating to see
that the payoff of the players is a monotonously
increasing function of amount of the entanglement,%
\begin{equation}
\$_{A}\left( i\sigma _{y},i\sigma _{y},i\sigma _{y}\right) =\$_{B}\left(
i\sigma _{y},i\sigma _{y},i\sigma _{y}\right) =\$_{C}\left( i\sigma
_{y},i\sigma _{y},i\sigma _{y}\right) =1+2\sin ^{2}\gamma .  \label{eq 6}
\end{equation}

\begin{figure}[t]
\centering{\includegraphics{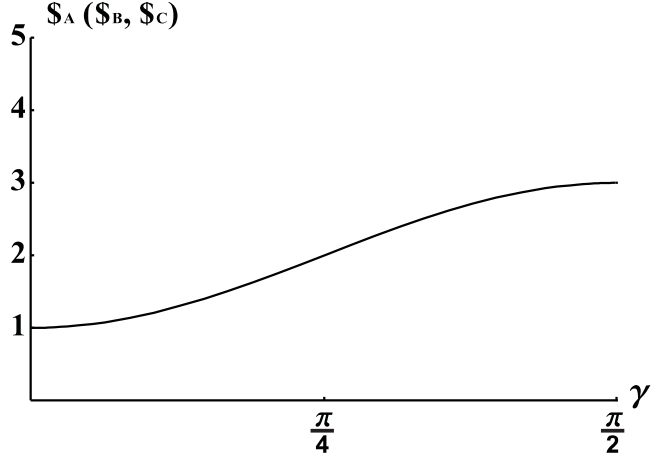}} \caption{The payoff plot as
a function of $\protect\gamma $ when all the
players resort to Nash Equilibrium, $i\protect\sigma _{y}\otimes i\protect%
\sigma _{y}\otimes i\protect\sigma _{y}$. From this figure we can
see that the payoffs of the players are the same and are
monotonous increasing function of $\protect\gamma $.} \label{Fig6}
\end{figure}

Fig \ref{Fig6} illustrates how the payoffs depend on the amount of
entanglement when the players all resort to Nash equilibrium. From
this figure, we can see that entanglement dominates the property
of the game: the payoffs of the players are the same, which is a
monotonous increasing function of the amount of entanglement of
the game's initial state. The profile $i\sigma _{y}\otimes i\sigma
_{y}\otimes i\sigma _{y}$ is always a Nash equilibrium of the game
independent of the entanglement, and the
dilemma is completely removed when the measure of game's entanglement $%
\gamma $ increases to its maximum $\pi /2$.

\subsubsection{General unitary operations}

In this section, we turn our attention to a more general situation, in which
players are allowed to adopt strategies from the whole unitary operations.
Just like in the two-player situation, the unitary operation can be denoted
as in Eq. (\ref{unitary operation}). Therefore a player's strategy can be
represented by a vector $\left( w,x,y,z\right) $.

At first, let us consider the case when the amount of entanglement is
maximal. We have known that in two player's Prisoner's Dilemma, there is no
pure strategy Nash equilibrium existing of the whole unitary space. However,
the situation is completely different in the three-player version of this
game. There indeed exist six symmetric Nash Equilibria strategy profiles $%
K_{1}\otimes K_{1}\otimes K_{1}$, $\cdots \cdots $, $K_{6}\otimes
K_{6}\otimes K_{6}$ with
\begin{eqnarray*}
K_{1} &=&\left( \frac{1}{\sqrt{2}},0,\frac{1}{\sqrt{2}},0\right) , \\
K_{2} &=&\left( \frac{1}{\sqrt{2}},0,-\frac{1}{\sqrt{2}},0\right) , \\
K_{3} &=&\left( -\frac{1}{2\sqrt{2}},\frac{\sqrt{3}}{2\sqrt{2}},\frac{1}{2%
\sqrt{2}},\frac{\sqrt{3}}{2\sqrt{2}}\right) , \\
K_{4} &=&\left( \frac{1}{2\sqrt{2}},-\frac{\sqrt{3}}{2\sqrt{2}},\frac{1}{2%
\sqrt{2}},\frac{\sqrt{3}}{2\sqrt{2}}\right) , \\
K_{5} &=&\left( \frac{1}{2\sqrt{2}},\frac{\sqrt{3}}{2\sqrt{2}},-\frac{1}{2%
\sqrt{2}},\frac{\sqrt{3}}{2\sqrt{2}}\right) , \\
K_{6} &=&\left( \frac{1}{2\sqrt{2}},\frac{\sqrt{3}}{2\sqrt{2}},\frac{1}{2%
\sqrt{2}},-\frac{\sqrt{3}}{2\sqrt{2}}\right) .
\end{eqnarray*}%
These six Nash Equilibria are symmetric to the players and yields the same
payoff $11/4$ to the three players. Hence, these Nash equilibrium keeps the
symmetry and fairness of the game, and are more efficient than the classical
Nash equilibrium $(D,D,D)$. In the following, we will take $K_{1}\otimes
K_{1}\otimes K_{1}$ as an example to prove that it is truly a Nash
equilibrium of the game. We write the strategy $K_{1}$ in matrix form%
\begin{eqnarray*}
K_{1} &=&\frac{1}{\sqrt{2}}\cdot I+0\cdot i\sigma _{x}+\frac{1}{\sqrt{2}}%
\cdot i\sigma _{y}+0\cdot i\sigma _{z} \\
&=&\left(
\begin{array}{cc}
\frac{1}{\sqrt{2}} & \frac{1}{\sqrt{2}} \\
-\frac{1}{\sqrt{2}} & \frac{1}{\sqrt{2}}%
\end{array}%
\right) .
\end{eqnarray*}%
Assume Bob and Colin both choose strategy $K_{1}$, according to payoff
function of Alice (see Eq. (\ref{threePDpayofffunction}))%
\begin{equation}
\$_{A}(U_{A},K_{1},K_{1})=\frac{1}{4}\left[ 11\left( w^{2}+y^{2}\right)
+10\left( x^{2}+z^{2}\right) \right] .  \label{eq 5}
\end{equation}%
From Eq. (\ref{eq 5}), we can see that $\$_{A}(U_{A},K_{1},K_{1})$ reaches
maximum if $w^{2}+y^{2}=1$. It is obviously that $w=y=1/\sqrt{2}$ satisfies
this condition. Therefore, Alice can get her best payoff when she chooses $%
K_{1}$ against the other two players' strategies $K_{1}\otimes K_{1}$.
Because the game is symmetric, the same analysis is true for Bob and Colin.
So $K_{1}\otimes K_{1}\otimes K_{1}$ is a Nash equilibrium of the game for
the whole set of unitary operations, which means that no player can increase
his individual payoff by unilateral deviating from the strategy profile.
From our proof, we can see that when both Bob and Colin adopt $%
K_{1}$, there exist other strategies that can yield the maximal payoff $%
11/4$ to Alice. Hence, $K_{1}\otimes K_{1}\otimes K_{1}$ is Nash
equilibrium, but not a strict one, as are the other five
equilibria.

\subsubsection{Non-maximal entanglement situation}

Unlike the situation of restricted strategic space, this time the
non-maximal entanglement situation is more complex and more
interesting. Since the solution we study here is symmetric and
fair to the players, we will take Alice as an instance.

\begin{figure}[t]
\centering{\includegraphics{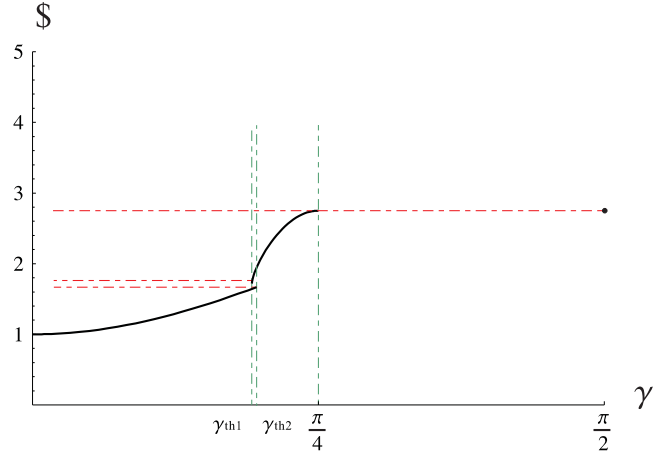}} \caption{The expected
payoff for Alice as a function of the measure of the parameter
$\protect\gamma $\ when all players resort to Nash Equilibrium
strategies.} \label{Fig7}
\end{figure}

Fig \ref{Fig7} depict Alice Payoff as a function of $\gamma $ when
the players all resort to Nash equilibrium strategy. From Fig
\ref{Fig7}, we see that the entanglement of the game is divided
into four domain by three thresholds, that are
\begin{eqnarray}
\gamma _{th1} &=&\arccos \sqrt{\frac{1}{26}\left( 13+\sqrt{\frac{3}{221}%
\left( 759+128\sqrt{42}\right) }\right) }\cong 0.60276,  \nonumber \\
\gamma _{th2} &=&\arcsin \sqrt{\frac{1}{3}}\cong 0.61548,  \nonumber \\
\gamma _{th3} &=&\frac{\pi }{4}\cong 0.78539.
\end{eqnarray}

In the domain $0\leqslant \gamma \leqslant \gamma _{th2}$, there exist four
Nash equilibria of the game, which are $\left( 0,0,\pm 1,0\right) $ and $%
\left( 0,\frac{\sqrt{3}}{2},\pm \frac{1}{2},0\right) $. These four Nash
equilibria yield the same payoff $1+2\sin ^{2}\gamma $. Hence, the player's
reward is a monotonous increasing function of $\gamma $, \textit{i.e.} the
game's property is enhance by the property of the game's entanglement. In
the domain $\gamma _{th1}\leqslant \gamma \leqslant \gamma _{th3}$, the
situation is very complex. The Nash equilibria of the game can be
represented as the following:%
\[
\pm \left[ a\left( \gamma \right) \left( 0,0,1,0\right) -b\left( \gamma
\right) \left( 0,0,0,\pm 1\right) \right]
\]%
and
\[
\pm \left[ a\left( \gamma \right) \left( \frac{\sqrt{3}}{2},0,0,\pm \frac{1}{%
2}\right) -b\left( \gamma \right) \left( 0,\frac{\sqrt{3}}{2},\pm \frac{1}{2}%
,0\right) \right] ,
\]%
where $a\left( \gamma \right) $, $b\left( \gamma \right) >0$ and
$a\left( \gamma \right) ^{2}+b\left( \gamma \right) ^{2}=1$. From
Fig \ref{Fig7}, we see that (i) the payoff in this domain is
bigger than the first domain and (ii) it also increases with the
increasing of entanglement. Hence, we can also get the conclusion
that in this domain, the game's property can be enhanced by the
entanglement of the game's state. In the domain $\gamma
_{th3}\leqslant \gamma <\pi /2$, there is no symmetric pure Nash
equilibrium. This situation is the same as the two-player version
of this game.

\section{Conclusion}

In this paper, we present the systematic investigation of quantum games with
the particular case of the Prisoner's Dilemma. By considering different
situations, the game shows properties which may outperform the
classical version of this game.

Quantum games and quantum strategies is a burgeoning field of quantum
information and quantum computation theory. Assuming the players
are playing the game by quantum rules, the game's solution is more
efficient than the classical one. Such quantum games are not just
esoteric exercises. They could form part of the longed-for quantum
technologies of tomorrow, such as ultra-fast quantum computers.
They might even help traders construct a crash-resistant stock
market. And quantum games could provide new insights into puzzling
natural phenomena such as high-temperature superconductivity
\textit{etc.}. Although at this stage, no one is sure which
applications will prove most fruitful, it is sure that quantum
game theory is a potential and promising research
field\cite{rf3}. Playing by quantum rules, every one will become a winner%
\cite{rf20}.

\section*{Acknowledgments}

We thank L. C. Kwek for carefully reading the paper. This project was supported by the National Nature Science
Foundation of China (Grants. No. 10075041 and No. 10075044) and
Funded by the National Fundamental Research Program (2001CB309300)
and the ASTAR Grant No. 012-104-0040.

\end{document}